\documentclass{aa}  
\usepackage{soul}
\usepackage{graphicx}
\usepackage{tabularx}
\usepackage{txfonts}
\usepackage{hyperref}

\begin{document}

\title{Connection-angle dependence of proton anisotropy in ground-level enhancement events}
  
\author{
    Alessandro~Bruno\inst{1,2}
    \and
    Silvia~Dalla\inst{3}
}

\institute{
Heliophysics Division, NASA Goddard Space Flight Center, Greenbelt, MD, USA\\
\email{alessandro.bruno-1@nasa.gov}
\and
Department of Physics, Catholic University of America, Washington, DC, USA
\and
Jeremiah Horrocks Institute, University of Lancashire, Preston PR1 2HE, UK
}

\authorrunning{Bruno et al.}
\titlerunning{Connection-angle dependence of proton anisotropy in GLE events}

\date{March 20, 2026. Accepted for publication in A\&A.}
  \abstract
  {
Ground Level Enhancements (GLEs) probe the earliest, highest-energy solar energetic particles and thus provide a unique window onto particle release and transport from the low corona to 1~AU. We present a uniform, event-resolved analysis of the early anisotropy for ten well-observed GLEs, combining consistently reconstructed neutron-monitor pitch-angle distributions (PADs) with Parker-spiral footpoint mapping. We find a clear, monotonic decline of initial anisotropy with increasing magnetic connection angle: well-connected events exhibit strong, persistent forward-directed beams, while poorly connected events show systematically weaker and more rapidly decaying anisotropies. This relationship holds across a wide range of flare classes and CME speeds, demonstrating that magnetic connectivity and interplanetary transport, rather than eruption magnitude, dominate the directional properties of the earliest relativistic arrivals at Earth. A principal component analysis was applied to time-resolved spectral and angular parameters to separate source-driven changes from transport effects. By explicitly identifying and removing secondary sunward (back-scattered) components—attributable to scattering and reflection from solar-wind structures and transient interplanetary features—from the PAD fits, we isolate the intrinsic relaxation of the primary forward beam and show that apparent departures from simple exponential decay are frequently attributable to reflected or delayed populations rather than prolonged source injection. 
The empirical anisotropy--connection-angle relation reported here provides an event-resolved, quantitative benchmark that constrains focused-transport and shock-acceleration models and offers immediate operational value: rapid footpoint estimates can meaningfully limit expected initial beaming and directional radiation risk.
}


\maketitle

\section{Introduction}
Ground Level Enhancements (GLEs) are the most extreme manifestations of solar energetic particle (SEP) activity, marked by (quasi-)relativistic protons that penetrate the atmosphere and produce statistically significant increases in ground-based neutron monitors (NMs) and muon detectors \citep{Simpson2000,Shea1990,Poluianov2025}. Because the highest-energy particles arrive at Earth within minutes of release, GLEs provide a uniquely sensitive probe of the timing, geometry, and early transport of particles accelerated in the low corona and at coronal shocks \citep{McCracken2008,Bieber2004}. The first-arriving population in many GLEs behaves as a narrowly collimated, nearly scatter-free beam, producing sharp, velocity-dispersed onsets and strong pitch-angle anisotropies consistent with near-ballistic propagation along the interplanetary magnetic field (IMF) and with adiabatic focusing in the diverging coronal and heliospheric field \citep{Skilling1971,Earl1976,Ruf1995,Ruffolo2006}. As events progress, scattering of particles in pitch angle, cross-field transport, and encounters with transient solar-wind structures -- interplanetary coronal mass ejections (ICME), magnetic clouds (MC) and their sheaths, interacting CMEs, and stream interaction regions (SIRs) -- broaden the distribution and reduce anisotropy \citep{Moraal2012,McCracken2008,ref:ROUILLARD2016}.
In some events, crossings of or proximity to the heliospheric current sheet (HCS) further modify the pitch‑angle distribution by altering magnetic connectivity and local scattering conditions \citep{Dalla2020,Waterfall2022}. Therefore, the anisotropy time history therefore encodes information about both the source release and the intervening transport environment.

\begin{table*}[!ht]
\centering
\caption{Relevant information for the ten GLE events analyzed in this study.}
\begin{tabular}{c|c|c|c|c|c|c|c}
\# & Event & Flare & Flare & CME & Solar-wind & Conn. & Bibliographic source\\
GLE & date & location & class & speed & state at onset & angle & for used NM data\\
\hline
45 & 1989 Oct 24 & S30W57 & X5.7 & 1497 & Disturbed$^{\dagger}$ & 24.5 & \citet{ref:GLE5-45}\\ 
59 & 2000 Jul 14 & N22W07 & X5.7 & 2061 & Inside ICME$^{a}$ & 36.2 & \citet{ref:GLE59-70}\\ 
60 & 2001 Apr 15 & S20W85 & X14.4 & 1199$^{*}$ & Inside ICME$^{a}$ & 41.5 & \citet{ref:GLE60} \\ 
65 & 2003 Oct 28 & S16E08 & X17 & 3128 & Trail of MC$^{a}$ & 42.5 & \citet{ref:GLE65-66} \\
66 & 2003 Oct 29 & S17N09 & X10 & 2628 & Trail of MC$^{a}$ & 29.6 & \citet{ref:GLE65-66} \\
67 & 2003 Nov 2 & S14W56 & X8.3 & 2733 & Trail of ICME$^{a}$ & 21.3 & \citet{ref:GLE67} \\
70 & 2006 Dec 13 & S06W23 & X3.4 & 2184 & Undisturbed$^{a}$ & 12.5 & \citet{ref:GLE59-70}\\
71 & 2012 May 17 & N11W76 & M5.1 & 1596 & Inside MC$^{a}$ & 16.9 & \citet{ref:GLE71}\\
72 & 2017 Sep 10 & S09W92 & X8.2 & 3163 & Trail of ICME$^{b}$ & 43.7 & \citet{ref:GLE72}\\
73 & 2021 Oct 28 & S28W01 & X1.0 & 1816 & Undisturbed & 77.7 & \citet{ref:GLE73}\\
\hline 
\end{tabular}
\\
\centering {\footnotesize 
$^{a}$\citet{ref:ROUILLARD2016}. $^{b}$\citet{ref:BRUNO2019}. 
$^{\dagger}$Uncertain (no data available). 
$^{*}$Partial halo CME; projected speed. 
}
\label{tab:GLEs}
\end{table*}

The worldwide network of NMs remains the principal observational tool for reconstructing the spectra and pitch-angle distributions (PADs) of the highest-rigidity ($\gtrsim$1 GV) SEPs \citep{Debrunner1994}. Spaceborne instruments such as 
the Payload for Antimatter Matter Exploration and Light-nuclei Astrophysics (PAMELA) detector
have provided direct, high-precision measurements of SEP energy spectra and PADs in the 80\,MeV to a few GeV range, anchoring NM yield-function calibrations and 
and reducing key systematic uncertainties
\citep{Bieber2004,Adriani2011,ref:ADRIANI2015,Bruno2018,ref:GLE71}. 
With stations distributed across a broad range of geomagnetic latitudes and longitudes, the NM network samples many asymptotic directions simultaneously, providing an effectively global ($\sim$4$\pi$) directional coverage that, together with their large effective area and continuous operation, makes NMs uniquely suited for tracking the spectral and angular evolution of the (near-)relativistic SEP population.

Despite extensive observational and modeling efforts, the principal controls on the magnitude and early evolution of anisotropy in GLEs are not fully settled. Several studies have sought correlations between anisotropy and eruption magnitude (flare class, CME speed) or shock properties, but published results are mixed, in part because of heterogeneous analysis methods, limited event samples, and uncertainties in magnetic mapping and NM response \citep[e.g.,][]{ref:CLIVER2020,ref:LARIOKARELITZ2014,ref:KOCHAROV2017}. Theoretical arguments and individual case studies have long suggested that magnetic connectivity—the angular separation between the SEP source at the Sun and the footpoint of the nominal Parker-spiral line passing through the observer—should exert a first-order control on early beaming \citep[e.g.,][]{RichardsonCane1996,McCracken2008}, yet a systematic, event-resolved quantification of this dependence across well-observed GLEs is lacking.

This paper addresses that gap by applying a uniform, event-by-event approach to the reconstruction and interpretation of early PADs in a curated set of GLEs. 
Our goal is to obtain consistent, comparable diagnostics of high-rigidity anisotropy and its temporal evolution, and to evaluate how these behaviors reflect underlying acceleration or transport conditions, with particular emphasis on their dependence on magnetic connection geometry.
To separate coupled spectral–angular evolution from independent transport effects, we apply principal component analysis (PCA) to the fitted spectral and PAD time series, identifying the dominant modes and testing whether anisotropy decay and spectral softening share a common origin.
In addition, we explicitly account for and separate secondary sunward (back-scattered) components -- associated with scattering and reflection from solar-wind structures and transient interplanetary features -- in the PAD fits so that the intrinsic behavior of the primary forward beam can be isolated for interpretation.

\section{NM observations}
\subsection{Data set}\label{Data set}
This study focuses on ten GLEs for which published NM reconstructions provide reliable, event-resolved rigidity spectra and PADs (Table~\ref{tab:GLEs}). 
The selected events span 1956--2021 and sample a wide range of flare classes, heliographic locations, and CME kinematics. For each event Table~\ref{tab:GLEs} lists the flare location (Column 3) and Geostationary Operational Environmental Satellite (GOES) soft X-ray class (Column 4), the CME three-dimensional (deprojected) speed (km s$^{-1}$) from the Coordinated Data Analysis Workshops (CDAW) halo catalog\footnote{\url{https://cdaw.gsfc.nasa.gov/CME_list/halo/}} (Column 5), and the heliospheric context at GLE onset (Column 6), indicating whether the Earth was inside a solar-wind structure, in its trailing region, or in nominal solar-wind conditions \citep{ref:ROUILLARD2016,ref:BRUNO2019}. No reliable solar-wind measurements are available for GLE~45.
However, several studies report highly-disturbed heliospheric conditions (see Section \ref{GLE45}). Column~7 gives the spherical magnetic connection angle between the flare site and the nominal Parker-spiral footpoint of IMF line passing through Earth (see Section~\ref{Connection Angle}). 
The final column lists the published NM reconstructions used in this work; we restrict the analysis to events reconstructed with the homogeneous University of Oulu inversion framework to ensure methodological consistency in the derived spectra and PADs across the sample.

\subsection{Spectral and angular reconstruction}
The reconstruction of primary rigidity spectra and PADs from global NM count-rate increases is carried out by forward-modeling parameterized spectral and angular forms through station-specific NM yield functions and geomagnetic transmission, and then fitting those model predictions to the observed network responses (e.g., \citealp{Mishev2020s}).  In practice the analysis adopts a separable ansatz for the differential intensity, \( J(R,\alpha) = J_{\parallel}(R)\,G(\alpha) \), where $R$ denotes particle rigidity, $\alpha$ the pitch angle measured relative to an instantaneous anisotropy axis, $J_{\parallel}$ the rigidity spectrum of the flux arriving from the Sun along the axis of symmetry, and $G$ the PAD (e.g., \citealp{ref:GLE71}).

Modern implementations perform the inversion in short successive time bins to recover time-dependent spectra and PADs. In the published Oulu NM reconstructions used here, the inversion is typically performed using 5-minute count-rate data. Because the anisotropy evolution examined in this work occurs on timescales of tens of minutes to hours, moderate changes in the integration interval (e.g., 10 minutes) would mainly smooth short-term fluctuations and are not expected to significantly affect the derived anisotropy trends.
Importantly, the NM yield function used in these reconstructions has been benchmarked against PAMELA and Alpha Magnetic Spectrometer-02 (AMS-02) high-quality space-borne measurements \citep{Bruno2018}, which provides an empirical cross-check of the NM response and reduces one important source of systematic uncertainty in the inferred high-rigidity spectra and angular distributions \citep{Koldobskiy2019,Mishev2020,ref:GLE71}.

\begin{figure*}
\centering
\caption{Temporal evolution (color code) of the PAD associated with the 10 GLEs analyzed in this study. }
\includegraphics[width=\linewidth]{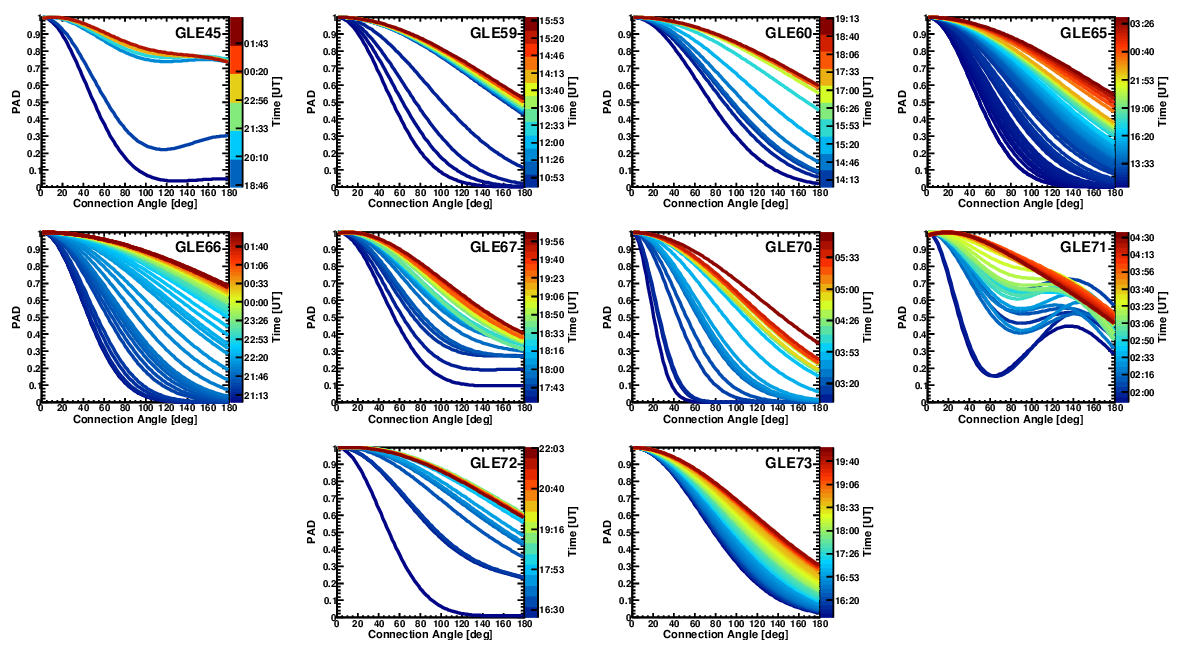}
\label{fig:PAD}
\end{figure*}

For the $>$1 GV rigidity spectrum, the NM analyses used in this study (see Table \ref{tab:GLEs}) adopt a modified power-law empirical form based on \citet{Cramp1997,Vashenyuk2006}:
\begin{equation}\label{eq:spectralshape}
J_{\parallel}(R) = J_{0}\,R^{-\left[\gamma + \delta\gamma (R-1)\right]},
\end{equation}
where $J_{0}$ is the normalization constant, $\gamma$ is the power-law spectral index at R=1~GV and $\delta\gamma$ is the rate of the spectrum steepening.
By enabling spectral softening as rigidity increases, the form removes the unphysical notion of a single unbroken power law extending to arbitrarily high energies, a feature likewise implicit in applications of the \citet{Band1993} double-power-law function (e.g., \citealp{TylkaDietrich2009}).
Nevertheless, Eq.~\ref{eq:spectralshape} is primarily a descriptive fit: its parameters ($J_{0}$, $\gamma$, $\delta\gamma$) are tuned to reproduce NM measurements but do not map uniquely onto microphysical quantities such as acceleration efficiency, mechanism‑specific spectral curvature, or a physically meaningful maximum rigidity. For inferring acceleration physics, physically motivated parameterizations—e.g., the \citet{EllisonRamaty1985} exponential‑cutoff—are highly preferable because they embed shock‑acceleration and escape assumptions and yield parameters that are more directly interpretable \citep{Bruno2018,Usoskin2020}.

The PAD is modeled with a single Gaussian function,
\( G(\alpha) = \exp\left(-\alpha^{2}/2\sigma^{2}\right) \),
where $\sigma$ is the distribution width,
defined with respect to an instantaneous anisotropy axis that is generally allowed to differ from the local IMF direction inferred from L1 spacecraft. This choice reflects the fact that, because of finite gyro-radius effects, the anisotropy axis is not necessarily aligned with the instantaneous IMF \citep{McCracken2008}. 
For simplicity, the PAD is further assumed to be independent of both the gyro-phase (azimuthal) angle and particle rigidity. 
Additional systematic uncertainties arise from the choice of older external geomagnetic field models \citep{Tsyganenko1989} versus more recent and accurate formulations \citep{ref:TSYGANENKO2005}, particularly under disturbed geomagnetic conditions.

\begin{figure*}[!t]
\centering
\caption{Dipole anisotropy (blue) and forward-inward anisotropy (red) as a function of time since event onset.}
\includegraphics[width=\linewidth]{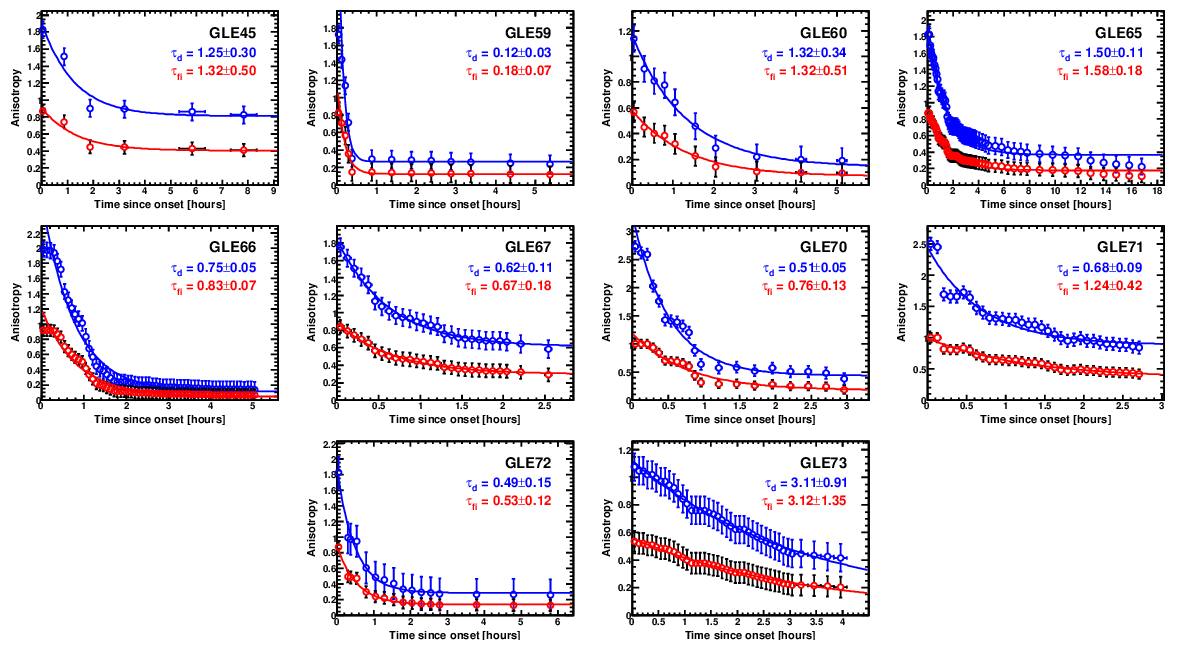}
\label{fig:Anisotropy_vs_time}
\end{figure*}

In the case of GLEs \#45, \#67, \#71 and \#72, a double-Gaussian functional form was used to account for a secondary peak associated with an early NM count-rate increases observed at several stations 
with anti-sunward asymptotic viewing direction 
\citep{ref:CRAMP1994,ref:GLE72,ref:GLE67,ref:GLE71,ref:GLE5-45}:
\begin{equation}\label{eq:doublepeak}
G(\alpha) \hspace{0.1cm} = \hspace{0.1cm} k \left[ \exp\left(-\alpha^{2}/2\sigma_{1}^{2}\right) + B \exp\left(-(\alpha-\alpha')^{2}/2\sigma_{2}^{2}\right) \right],
\end{equation}
where $k$ is the normalization constant so that $0\leq G(\alpha)\leq1$, $\sigma_{1}$ and $\sigma_{2}$ are the width parameters of the PAD, and $B$ and $\alpha'$ (typically = $\pi$) define the back-scattered (inward) component.
This feature has been typically interpreted as an effect of scattering off particle-generated turbulence within 1 AU, or
disturbed interplanetary transport conditions associated with transient solar-wind structures. Such structures not only cause deviations from the nominal IMF geometry (see, e.g., \citealp{ref:RICHARDSON1991,RichardsonCane1996}), but also tend to to decrease or increase the intensity of SEP events, depending on their relative location with respect to the Earth \citep{ref:LARIOKARELITZ2014}.

\section{Anisotropy evaluation}
To quantify the anisotropy of each GLE, we evaluated the time‑dependent SEP angular fluxes in the forward (anti‑sunward) and inward (sunward) hemispheres, \(I_{\mathrm{fw}}(t)\) and \(I_{\mathrm{in}}(t)\), obtained by integrating \(G\) over the pitch‑angle intervals \([0,\pi/2]\) and \([\pi/2,\pi]\), respectively:
\begin{equation}
I_{\mathrm{fw}}(t) = \int_{0}^{\pi/2} G(\alpha,t)\,\sin\alpha\, d\alpha ,
\qquad
I_{\mathrm{in}}(t) = \int_{\pi/2}^{\pi} G(\alpha,t)\,\sin\alpha\, d\alpha .
\end{equation}
The forward–inward anisotropy is then computed as
\begin{equation}\label{anisotropy}
A_\mathit{f/i} (t) = \frac{I_{\mathit{fw}} (t)-I_{\mathit{in}} (t)}{I_{\mathit{fw}} (t)+I_{\mathit{in}} (t)},
\end{equation} 
varying between -1 and +1.
Similarly, following \citet{Bieber2002}, an instantaneous dipole anisotropy can be estimated as:
\begin{equation}\label{dipole_anisotropy}
A_{d} (t) = 3\hspace{1mm}\frac{\int_{-1}^{+1}G(\alpha,t) \, \cos\alpha \, d\cos\alpha}{\int_{-1}^{+1}G(\alpha,t) \, d\cos\alpha},
\end{equation} 
varying between -3 and +3.
The two anisotropy measures defined above provide complementary information on the angular structure of the SEP distribution. The forward--inward anisotropy $A_{\mathrm{f/i}}(t)$ offers a direct geometric comparison between the anti-sunward and sunward hemispheres. Because it depends only on the integrated fluxes $I_{\mathrm{fw}}$ and $I_{\mathrm{in}}$, this quantity is relatively insensitive to uncertainties in the detailed shape of the PAD and remains stable even when the distribution departs from a simple dipolar form. Values approaching $+1$ indicate a strongly forward-directed beam, values near zero correspond to near-isotropy, and negative values reflect sunward dominance. In parallel, the dipole anisotropy $A_{d}(t)$ represents the first moment of the PAD and is directly interpretable within focused-transport theory. While $A_{d}(t)$ provides a physically meaningful measure that can be compared across instruments and events, it is more sensitive to uncertainties in the PAD reconstruction. Using both metrics together therefore combines the robustness and geometric transparency of $A_{\mathrm{f/i}}(t)$ with the physical interpretability of the dipole moment, yielding a consistent characterization of the anisotropy throughout the event.

The temporal evolution of the anisotropy is commonly described by an exponential decay,
\begin{equation}
A(t) \simeq A_{0}\,\exp\!\left(-\frac{t-t_{0}}{\tau}\right),
\end{equation}
which represents the gradual 
relaxation of an initially field-aligned beam under particle scattering that increases the spread of pitch angles. In focused-transport theory, such a form naturally arises when a single population of forward-streaming particles undergoes diffusive isotropization with a characteristic timescale $\tau$, leading to a monotonic decrease of the anisotropy that is well captured by a single exponential function \citep{Skilling1971,Earl1976,Ruffolo2006,Schlickeiser2002,Marsh2015}.

When analyzing GLEs with substantial back-scattered components (\#45, \#67, \#71, \#72), single exponentials frequently fail because the observed PAD is the time-dependent superposition of at least two components: a prompt, narrow forward beam and a delayed, broader or sunward population. This superposition produces non-monotonic or multi-stage anisotropy profiles that cannot be captured by a single characteristic timescale, yielding poor exponential fits. Conversely, setting the back-scattered term to zero in Eq.~\ref{eq:doublepeak} ($B=0$) and fitting only the forward-streaming component restores a monotonic relaxation and yields stable, physically meaningful exponential fits. Thus the departures from exponential behavior reflect the time-dependent admixture of back-scattered flux rather than an intrinsic non-exponential evolution of the primary beam. 

The resulting time profiles of $A_{\mathrm{f/i}}$ and $A_{d}$—after removing the back‑scattered component, when present—for the ten GLE events are shown in Figure~\ref{fig:Anisotropy_vs_time}. The rapid, highly anisotropic onset phase is generally attributed to relativistic protons released in the low corona, where strong adiabatic focusing produces a narrow, field–aligned beam that propagates to 1~AU with very little scattering \citep{McCracken2008}. As the event progresses, particle scattering by magnetic inhomogeneities and Alfv\'en waves in the interplanetary magnetic field redistributes their pitch angles and gradually isotropizes the distribution, producing a delayed, more broadly distributed
population. The transition between these two regimes---from a prompt, forward–directed beam to a more isotropic flux---can be explained by the late arrival of particles that have been
scattered to large pitch angles during interplanetary transport \citep{ref:LI2019}. The isotropization times $\tau_{\mathrm{f/i}}$ and $\tau_{d}$ extracted for each event from both anisotropy definitions  are shown in its respective panel, providing a quantitative measure of the rate at which the anisotropy relaxes.

\section{Event interpretation and PCA diagnostics}
The following subsections examine the individual GLE, linking fitted PAD morphology and anisotropy time series (Figures~\ref{fig:PAD}--\ref{fig:Anisotropy_vs_time}) to source properties, Parker-spiral footpoints, and heliospheric context (Table~\ref{tab:GLEs}). We summarize PAD evolution, report isotropization diagnostics (forward and full PAD), and synthesize these with local solar-wind state to infer the dominant transport processes. Each subsection also presents PCA of the spectral \((J_0,\gamma,\delta\gamma)\) and PAD — \(\sigma_{1}(t)\) and, when present, \(\sigma_{2}(t),\,B(t),\,\alpha'(t)\) — time series. 
PCA is used to identify the dominant, independent patterns in a multivariate dataset and ranks them by the fraction of variance they explain. The mode amplitudes show how those patterns evolve in time, and the mode loadings indicate which measured quantities (e.g., spectral hardness or PAD metrics) contribute most to each pattern, allowing us to distinguish coupled spectral–angular behavior from independent angular or spectral modulation. We therefore use PCA to separate source-driven changes from transport effects and to determine whether spectral softening and anisotropy decay reflect a single underlying trend or evolve independently. PCA results are compared across events in Section~\ref{Comparative Synthesis of PCA Results}.

\subsection{GLE 45 (1989 Oct 24)}\label{GLE45}
The 24 October 1989 event (X5.7; S30W57) was associated with a fast CME ($\sim$1497 km s$^{-1}$) and occurred amid an intense sequence of preceding eruptions that left the local heliosphere and magnetosphere highly disturbed. While direct solar-wind measurements are lacking (see Section \ref{Data set}), observational studies document perturbed geomagnetic conditions, and suggest altered IMF topology and remote scattering/reflecting solar-wind structures that increase uncertainty in magnetic mapping and station-to-station timing \citep{Reeves1992,ref:CRAMP1994,McCracken2008,Shea1990,ref:GLE5-45}. 
 
Reconstructed PADs show a strong forward peak at onset that fills in and becomes bidirectional within minutes; fitted isotropization times are \(\tau_d=1.25\pm0.30\ \mathrm{h}\) (dipole) and \(\tau_{fh}=1.32\pm0.50\ \mathrm{h}\). These comparable, relatively long decay constants are best interpreted as the convolution of (1) an initially well-focused low-coronal injection, (2) contamination of the forward beam by reflected/mirrored flux and magnetospheric reprocessing that sustains backward flux, and (3) enhanced pitch-angle scattering from transient IMF distortions and sheath/turbulence that progressively isotropize the distribution. Consequently, the measured \(\tau\) values should be treated as effective isotropization times that combine interplanetary scattering and magnetospheric/mirroring contributions; neglecting remote reflection or cutoff corrections will bias transport inversions toward artificially long parallel mean free paths and misattribute PAD filling to source duration rather than to propagation and magnetospheric effects \citep{Reeves1992,ref:CRAMP1994,Shea1990,ref:GLE5-45}. 

GLE~45 exhibits a tightly coupled evolution of particle spectrum and anisotropy, and the PCA provides quantitative support for a transport-dominated interpretation. The first principal component links an initially hard, intense, and strongly beamed forward population to a progressive spectral softening and concurrent PAD broadening, indicating that increasing scattering and weakening magnetic connection drive both spectral decay and loss of anisotropy. A secondary component, which accounts for a smaller fraction of the variance, isolates an anti-sunward contribution that grows steadily as the main beam weakens. Together these results indicate two physical populations: a rapidly evolving, well-connected forward beam whose hardness and collimation vary in tandem, and a slower, back-scattered component that becomes more prominent as scattering intensifies.

\subsection{GLE 59 (2000 Jul 14)}
The Bastille Day eruption (N22W07, X5.7) produced a very fast CME (\(\sim2061\ \mathrm{km\,s^{-1}}\)); coronagraph and radio diagnostics place shock formation low in the corona and show wide longitudinal shock signatures that enabled broad particle access \citep{Belov2001,Duldig2001,Bieber2002,Vashenyuk2006,Wu2020}. NM inversions and time-resolved modeling report a hard, prompt relativistic component at onset followed by extended emission and rapidly evolving anisotropy \citep{Belov2001,Bieber2002}. 

Interplanetary reconstructions place Earth inside ICME material at onset, implying field rotations, compressed sheaths, and enhanced turbulence that modify Parker-spiral mapping and promote rapid scattering and cross-field access \citep{ref:ROUILLARD2016}. The PAD fits show an extremely sharp forward peak with very short isotropization times, \(\tau_d=0.12\pm0.03\ \mathrm{h}\) and \(\tau_{fh}=0.18\pm0.07\ \mathrm{h}\), indicating the impulsive beam is erased on minute timescales. The most plausible interpretation is an impulsive low-coronal injection rapidly supplemented by broad shock-mediated injections across many flux tubes, with strong local turbulence in the shock/sheath driving fast isotropization \citep{Bieber2002,Vashenyuk2006}.

GLE~59 behaves as a remarkably 
clean, single-mode event in which the spectrum and anisotropy evolve in lockstep. The first component contains more than 90\% of the variance, meaning that the hardening/softening of the spectrum and the narrowing/broadening of the PAD follow the same underlying physical trend. Early in the event the spectrum is relatively hard and the anisotropy strong, consistent with a well-connected, forward-moving relativistic beam. As time progresses, the spectrum softens steadily while the PAD broadens, reflecting increasing scattering and a gradual loss of magnetic connection. The higher-order components contribute very little, indicating that GLE~59 does not contain a significant secondary or back-scattered population: its evolution is dominated by a single, coherent process linking spectral decay to anisotropy decay. This makes GLE~59 one of the clearest examples where the temporal behavior of the spectrum directly mirrors the evolution of the anisotropy, pointing to transport effects as the primary driver of the event’s time profile.

\subsection{GLE 60 (2001 Apr 15)}
The S20W85 (X14.4) eruption produced a fast, partial-halo CME (\(\sim1199\ \mathrm{km\,s^{-1}}\)) and a rapid, strongly anisotropic proton onset consistent with good nominal magnetic connectivity. Modern NM reanalyses provide time-dependent rigidity spectra and angular distributions for this event \citep{ref:GLE60}, while contemporaneous solar-wind and reconstruction studies identify disturbed IMF intervals and transient structures (including ICME/sheath material) that modify simple Parker-spiral footpoint mapping \citep{Gopalswamy2005,ref:ROUILLARD2016}. 

The PADs show a canonical evolution from a narrow, forward-peaked beam to a broader, hour-scale relaxation. Measured isotropization times are \(\tau_d=1.32\pm0.34\ \mathrm{h}\) and \(\tau_{fh}=1.32\pm0.51\ \mathrm{h}\), indicating a persistent anisotropy that decays on \(\sim\)hour timescales. The data are most consistent with a combination of sustained/extended injection (e.g., sampling a shock flank) and moderate early scattering (finite parallel mean free path), with later encounters with transient solar-wind structures accelerating isotropization. 

GLE~60 shows a clear, physically coherent pattern in which the spectrum and anisotropy evolve together, but with a more gradual transition than in the very clean single-mode events. The dominant component 
($>$80\% variance)
reflects the familiar trend: the event begins with a relatively hard spectrum and a narrow PAD, indicating a well-focused forward beam and good magnetic connection. As the event progresses, the spectrum softens steadily while the PAD broadens, showing that increasing scattering and weakening connection drive both the spectral decay and the loss of anisotropy. The second component
($\sim$18\% variance)
reflects additional structure in the anisotropy evolution—particularly the late-time broadening and the rapid decline of the directional signal once the spectrum becomes soft. This indicates that GLE~60 is still dominated by a single transport-controlled mode, but with a secondary contribution that modulates the anisotropy more strongly than the spectrum. Overall, the PCA confirms that the hard-to-soft spectral evolution and the narrowing-to-broadening PAD evolution are tightly linked, with transport effects shaping the event from start to finish.

\subsection{GLE 65 \& GLE 66 (2003 Oct 28--29)}\label{GLE65-66}

The 28–29 October 2003 eruptions (28 Oct: S16E08, X17; 29 Oct: S17W09, X10) produced exceptionally fast CMEs (3128 and 2628\,km\,s\(^{-1}\)) whose shocks formed low in the corona and propagated into a highly structured inner heliosphere. Coronagraph, radio, and imaging studies document early shock formation and CME–streamer interactions that enhance shock strength and particle injection efficiency \citep{Gopalswamy2005}. Global NM inversions and transport analyses indicate merged or closely spaced shocks, compressed sheaths, and elevated turbulence that substantially modified Parker-spiral connectivity to Earth \citep{Gopalswamy2005JGRA,Vashenyuk2006}; interplanetary reconstructions place Earth in the wake of a magnetic cloud during the onsets \citep{ref:ROUILLARD2016}.

The PADs are among the most structurally complex in our sample. For GLE\,65 we measure \(\tau_d=1.50\pm0.11\ \mathrm{h}\) and \(\tau_{fh}=1.58\pm0.18\ \mathrm{h}\); for GLE\,66 the decay constants are shorter (\(\tau_d=0.75\pm0.05\ \mathrm{h}\), \(\tau_{fh}=0.83\pm0.07\ \mathrm{h}\)). Both events show very narrow forward peaks at onset that rapidly evolve into multi-component or bidirectional shapes as merged shocks and compressed sheaths introduce additional injection sites and strong turbulence. The longer persistence of anisotropy in GLE\,65 versus the faster isotropization in GLE\,66 likely reflects differences in local shock geometry, sheath encounter timing/strength, and the degree of CME–CME interaction along Earth-connected field lines.

For GLE\,65 the multivariate evolution is substantially more complex than in the earlier single-mode events. The analysis shows that the spectrum and anisotropy do not follow a single, tightly coupled trend: a primary pattern ($\sim$66\% of the variance) reproduces the familiar transport-driven link between spectral softening and PAD broadening, while a secondary pattern ($\sim$33\% of the variance) captures a slower, large-amplitude evolution of the angular distribution that is not reflected in the spectral index. In the time series this appears as a smoothly softening spectrum concurrent with a much more abrupt collapse of anisotropy, the PAD width increasing from narrow early values to very broad late-time values even when the spectrum changes only modestly. Physically, this behavior implies that GLE\,65 cannot be described by a single transport-controlled mode: the event evolves from an initially well-collimated beam to a phase dominated by strong scattering and redistribution, in which angular redistribution proceeds faster and with greater amplitude than spectral softening.

GLE\,66 exhibits a related two-mode evolution but with different timing and relative strengths. At onset the spectrum and PAD evolve together, consistent with a well-connected, shock-driven injection; at later times a pronounced secondary pattern ($>$25\% of the variance) emerges and captures a rapid collapse of anisotropy that is not matched by an equivalent spectral change. This behavior indicates a transition from a transport-controlled phase to one in which angular redistribution (enhanced scattering, sudden loss of magnetic connection, or local sheath effects) dominates the observed evolution: the PAD broadens and the directional signal decays more rapidly than the integrated spectrum softens, producing a genuine decoupling between spectral and angular evolution.

\subsection{GLE 67 (2003 Nov 2)}
The S14W56 (X8.3) eruption produced a very fast CME (\(\sim2700\ \mathrm{km\,s^{-1}}\)); timing analyses indicate the relativistic release began while the CME nose remained low in the corona, and multi-wavelength diagnostics (continuum emission and late, low-frequency type~III onsets) support an early, flare/CME-launch acceleration phase \citep{ref:KOCHAROV2017}. At the same time Earth lay within the disturbed Halloween heliosphere, with residual ICMEs and interacting transients that substantially altered magnetic connectivity \citep{ref:ROUILLARD2016}. 

The PADs show an initially narrow forward beam that broadens rapidly; fitted isotropization times are \(\tau_d=0.62\pm0.11\ \mathrm{h}\) and \(\tau_{fh}=0.67\pm0.18\ \mathrm{h}\), indicating efficient early scattering. This combination—prompt low-coronal release but fast angular filling—implies a strongly focused source whose beam is rapidly redistributed by enhanced diffusion of particles in pitch angles and cross-field mixing in the disturbed heliospheric environment; the early beam width and first-moment amplitudes therefore place upper limits on the parallel mean free path. 

GLE~67 is a classic two-component event where the forward beam and the back-scattered population evolve on different timescales. The main component shows the expected link between spectral softening and PAD broadening, but the second component ($\sim$30\% variance) captures the independent growth of the anti-sunward peak. As the event progresses, the forward beam weakens and becomes more isotropic, while the back-scattered component strengthens and persists, even when the spectral index changes only modestly. This produces a clear decoupling: the spectrum evolves smoothly, but the anisotropy undergoes a more complex, two-stage evolution driven by both transport and reflection. GLE~67 therefore exhibits a richer angular structure than single-mode events, with the PCA cleanly separating the forward-beam decay from the buildup of the back-scattered population.

\subsection{GLE 70 (2006 Dec 13)}
The S06W23 (X3.4) eruption produced a fast CME (\(\sim2184\ \mathrm{km\,s^{-1}}\)) and occurred under comparatively undisturbed solar-wind conditions with near-nominal Parker-spiral connectivity \citep{ref:ROUILLARD2016}. Early radio and coronagraph timing support prompt shock formation low in the corona and rapid field-aligned injection from the western source, consistent with the prompt, strongly anisotropic NM onsets \citep{Adriani2011,ref:GLE59-70}. 

PAMELA’s high-precision in-situ spectra anchor the low-to-mid energy portion of the SEP population ($\sim$80 MeV/n to $\sim$3 GeV/n), fixing absolute normalization and revealing spectral features that, together with NM fits, constrain acceleration efficiency and transport (scattering and possible cross-field spreading) from low coronal heights \citep{Adriani2011,ref:GLE59-70}. The PADs are clean and strongly forward-peaked at onset; fitted decay times are \(\tau_d=0.51\pm0.05\ \mathrm{h}\) and \(\tau_{fh}=0.76\pm0.13\ \mathrm{h}\), implying rapid loss of small-angle structure while the net forward excess persists slightly longer. 

GLE~70 behaves as a clean, single-mode event in which spectral and anisotropy evolution remain tightly coupled. The first component explains 90\% of the variance, showing that the spectrum hardens and the PAD narrows together early on, then soften and broaden in parallel as scattering increases. The higher components contribute very little and mainly reflect small fluctuations in the PAD width. Physically, this means GLE~70 is dominated by a single forward-moving beam whose intensity, hardness, and collimation rise and fall coherently. There is no evidence for a significant secondary or back-scattered population. The event’s evolution is therefore governed almost entirely by transport effects acting on a single, well-connected beam.

\subsection{GLE 71 (2012 May 17)}
The eruption (N11W76, M5.1; CME $\sim$1596\,km\,s$^{-1}$) produced a prompt, field-aligned relativistic injection at nominally well-connected longitudes, but the heliospheric context was unusually complex. Earth was embedded in the magnetic cloud and sheath of a prior CME, and the coronal/shock geometry evolved rapidly during the release interval. Coronal mapping shows that the expanding shock intersected different footpoints at different times, while CME deflection, rotation, and interchange reconnection generated mixed flux-tube connectivity and locally altered field orientations \citep{ref:ROUILLARD2016,ref:PALMERIO2021}. Together with NM inversions and PAMELA observations, these constraints establish GLE\,71 as a case where an efficient low-coronal accelerator produced a prompt, hard relativistic component whose signatures were rapidly modified by the disturbed propagation environment \citep{ref:ADRIANI2015,Bruno2016,Bruno2018,ref:GLE71}.

The PADs reflect this joint source–transport picture. The global NM network shows a brief ($\sim$5\,min) interval of {\em negative} forward–inward anisotropy, indicating transient anti-sunward dominance caused by evolving shock–field geometry and local field rotations \citep{ref:GLE71,ref:ROUILLARD2016}. PAMELA provides the rigidity-resolved counterpart: at $R\gtrsim1$\,GV both datasets show a sharply collimated, field-aligned beam, while a weaker enhancement at $\alpha\approx135^\circ$–$140^\circ$ reflects partial reflection or mirroring within the distorted ICME flux rope or return of particles scattered beyond 1\,AU \citep{ref:PALMERIO2021,ref:ROUILLARD2016}.

During the first tens of minutes the anisotropy axis remains aligned with the IMF and the PAD is dominated by the forward beam. As the shock connects to additional flux tubes and scattering increases in the ICME/sheath, the PAD broadens and the sunward 
component strengthens. This evolution parallels the drift of the NM-reconstructed source direction and the chronology of shock–field intersections \citep{ref:ROUILLARD2016}. Spectrally, PAMELA observes an early hard component extending into the GeV range followed by softening concurrent with angular broadening, consistent with scatter-free propagation of the highest-rigidity particles and rapid redistribution of lower-rigidity ones \citep{ref:ADRIANI2015,Bruno2016,Bruno2018}.

The dipole anisotropy decays with $\tau_d=0.68\pm0.09$\,h, while the forward–inward index decays more slowly, $\tau_{fh}=1.24\pm0.42$\,h, both estimated excluding the back-scattered component. The divergence reflects rapid loss of coherence in the narrow high-rigidity beam versus a more persistent hemispheric asymmetry maintained by continued injections and magnetic mirroring in compressed sheath regions.

GLE\,71 is the most complex event in the set, with several independent modes contributing comparable variance. The leading mode captures the familiar transport-driven link between spectral softening and PAD broadening, but three additional modes (together accounting for $\sim$40\% of the variance) introduce independent structure in both the forward and backward peaks. The back-scattered population evolves on its own timescale, and the widths of the two Gaussian components change differently as the event progresses. The resulting evolution is intrinsically multi-dimensional: the spectrum softens steadily while the angular distribution undergoes distinct phases (including a mid-event tightening and a late-time broadening) that are not mirrored in the spectral index. In this sense GLE\,71 is best described as a genuinely complex event, where multiple comparable modes (transport, reflection/mirroring, and redistribution) act simultaneously and preclude reduction to a single physical trend.

\subsection{GLE 72 (2017 Sep 10)}
The parent eruption (S09W92, X8.2) launched an exceptionally fast CME ($\sim$3160\,km\,s$^{-1}$) while Earth was inside the trailing edge of a prior ICME and in the recovery phase of a Forbush decrease \citep{ref:BRUNO2019}. These disturbed conditions altered nominal Parker-spiral connectivity and elevated background turbulence. The ground-level response showed a rapid, strongly field-aligned rise and a hard high-energy spectrum, but with pronounced early temporal structure—low-energy breaks and a high-energy rollover—that evolved on minute-to-hour timescales. Together with prompt radio and white-light shock signatures, these features indicate highly efficient low-coronal shock acceleration, while the disturbed heliosphere implies that longitude-based connectivity alone cannot explain the earliest fluxes \citep{ref:BRUNO2019,ref:GLE72}.

PADs exhibit an initially narrow forward peak at well-connected longitudes that quickly broadens as ICME/sheath turbulence acts. The short, nearly equal isotropization times, $\tau_d=0.49\pm0.15$\,h and $\tau_{fh}=0.53\pm0.12$\,h, point to rapid, rigidity-dependent scattering and fast cross-field spreading that erase the initial connectivity imprint.

GLE~72 shows a dominant single-mode evolution early on, followed by a strong secondary contribution associated with the back-scattered population. The first component reflects the expected hard-to-soft spectral evolution and narrowing-to-broadening PAD trend. But the second and third components capture the rapid growth of the anti-sunward peak and the late-time collapse of the anisotropy, even when the spectrum changes only modestly. This indicates that GLE~72 transitions from a clean forward-beam phase to one dominated by strong scattering and reflection, with the back-scattered component becoming increasingly important. The PCA therefore reveals a two-stage event: an early transport-controlled phase and a later phase shaped by redistribution and bidirectional structure.

\subsection{GLE 73 (2021 Oct 28)}
The eruption (S26W05, X1.0) launched a southward-directed CME that weakened nominal Sun–Earth magnetic linkage, contributing to the delayed relativistic onsets at many stations. Multiwavelength timing places type-III/II/IV activity at 15:28–15:30\,UT, an EUV wave at $\approx$15:39\,UT, and a $\ge$1\,GV release at $\approx$15:40\,UT when the white-light shock was at $\approx 2.3\,R_{\odot}$. These constraints support prompt low-coronal acceleration while allowing for additional, spatially distributed injections that could broaden the PAD at 1\,AU. NM reconstructions show a moderately hard initial spectrum with a high-rigidity roll-off and an unusually broad PAD \citep{Struminsky2023,ref:GLE73,Papaioannou2022}.

PADs are broad from the outset, consistent with weakened field-aligned access and substantial cross-field spreading or mixed flux-tube connectivity. The long isotropization times, $\tau_d = 3.11\pm0.91$\,h and $\tau_{fh} = 3.12\pm1.35$\,h, indicate slow anisotropy decay and favor an interpretation involving spatially and temporally distributed injections and prolonged lateral access rather than a single weakly scattered beam—consistent with the southerly source, likely CME deflection, and NM evidence for broad PADs.

PCA reinforces this view. GLE~73 is dominated by a single coherent mode linking spectral softening to anisotropy decay, but with a noticeable secondary contribution that modulates the PAD more strongly than the spectrum. Early in the event, the spectrum is moderately hard and the PAD relatively narrow; both evolve smoothly as scattering increases. The second component reflects a slower, independent broadening of the PAD that becomes more important late in the event, even when the spectral index changes only gradually. This produces a mild decoupling: the spectrum evolves steadily, but the anisotropy collapses more rapidly. Overall, GLE~73 remains largely single-mode, but with a secondary angular trend that becomes visible as the event approaches its isotropic tail.

\subsection{Comparative synthesis of PCA results}\label{Comparative Synthesis of PCA Results}
The PCA diagnostics are reported in Table~\ref{tab:pca_summary}. \emph{PC1 var} and \emph{PC2 var} are the fractions of the total variance explained by the first and second principal components, respectively; they quantify how much of the joint spectral–angular variability is captured by each mode. To compactly characterize event complexity we introduce the effective dimensionality
$D_{\mathrm{eff}} \;=\; 1 / \sum_{i} \lambda_i^2$, with $\sum_i \lambda_i$ = 1,
where $\lambda_i$ are the PCA eigenvalue fractions normalized to unit sum. Values of $D_{\mathrm{eff}}$ near unity indicate that a single coupled spectral–angular mode dominates the evolution, while larger values signal the presence of multiple, independent processes (for example, a prompt beam plus a delayed reflected or redistributed population). 

\begin{table*}
\centering
\caption{PCA diagnostics for the 10 GLEs.}
\footnotesize
\setlength{\extrarowheight}{2pt}
\begin{tabular}{c|c|c|c|c|l}
GLE & PC1 var (\%) & PC2 var (\%) & $D_{\rm eff}$ & Classification & Interpretation \\
\hline
45  & 88.1 & 11.8 & 1.27 & Multi-component & Prompt hard forward beam; delayed\\
    &      &      &      &                 & back-scatter from disturbed heliosphere. \\
59  & 91.1 & 7.2  & 1.20 & Single-mode      & Strong prompt beam; minimal \\
    &      &      &      &                 & transport/reflection signature. \\
60  & 81.3 & 18.1 & 1.44 & Single-mode      & Coupled spectral–angular evolution;\\
    &      &      &      &                 & moderate transport modulation. \\
65  & 66.7 & 30.7 & 1.85 & Multi-component & Significant secondary/transport\\
    &      &      &      &                 & component; multi-stage evolution. \\
66  & 71.4 & 26.0 & 1.73 & Multi-component & Primary beam present; notable \\
    &      &      &      &                 & transport/reflection effects. \\
67  & 69.2 & 27.9 & 1.80 & Multi-component & Prompt beam with back-scattered \\
    &      &      &      &                 & population linked to ICME trail. \\
70  & 90.0 & 6.8  & 1.23 & Single-mode      & Very strong single-mode prompt\\
    &      &      &      &                 &  beam; negligible secondary mode. \\
71  & 58.7 & 16.0 & 2.52 & Complex         & Weak PC1 dominance; complex, multi-\\
    &      &      &      &                 & component transport and reflections. \\
72  & 77.7 & 15.2 & 1.59 & Multi-component & Primary forward beam with a secondary\\
    &      &      &      &                 & reflected/back-scattered component. \\
73  & 79.6 & 18.7 & 1.49 & Multi-component & Moderate single-mode behavior; \\
    &      &      &      &                 & transport signatures present. \\
\hline
\end{tabular}
\label{tab:pca_summary}
\end{table*}

Taken together, the PCA results show that GLEs in the sample occupy a continuum between two limiting behaviors defined by the relative importance of a single coupled mode and additional transport-driven components.

A first group—GLE~59, 60, and 70—is dominated by a single physical mode in which the spectrum and anisotropy evolve in lockstep. These events begin with a hard, narrowly beamed forward population that softens and broadens smoothly as scattering increases, with no significant secondary component. Their hardness–anisotropy evolution is therefore a clean tracer of transport and magnetic connection, with minimal contamination from reflection or redistribution.

A second group—GLE~45, 65, 66, 67, 72, and 73—shows clear evidence for additional structure beyond the primary transport-controlled mode. In these events, the anisotropy evolves on at least two timescales. For GLE~45, 67, and 72, this reflects the presence of a genuine back-scattered (anti-sunward) population that grows and persists independently of the forward beam. For GLE~65, 66, and 73, there is no distinct back-peak, but the PAD still undergoes a stronger and more complex evolution than the spectrum: the anisotropy collapses or redistributes more rapidly than $\gamma(t)$ changes. In all these cases, hardness vs anisotropy still carries information about transport, but it no longer captures the full physics of the event, because part of the angular evolution is driven by processes—reflection or redistribution—that do not have a one-to-one spectral counterpart.

GLE~71 is an extreme example of the multi-component regime, with several modes contributing comparable variance and with forward and backward peaks, widths, and relative amplitudes evolving on distinct timescales. Here the spectral and angular evolutions decouple: spectral softening proceeds while the PAD undergoes complex, multi-phase restructuring driven by a combination of transport, reflection, and redistribution processes.

Across the cycle, the PCA therefore reveals a continuum: from clean, single-mode events where spectral hardness and anisotropy are inseparable, to multi-component events where angular structure evolves independently of the spectrum. This classification provides a physically grounded framework for interpreting hardness–anisotropy correlations and for distinguishing events dominated by simple transport from those shaped by more complex heliospheric processes.

\section{Connection-angle dependence of GLE anisotropy}

\subsection{Connection Angle}\label{Connection Angle}
The magnetic connection angle is a key parameter controlling relativistic-particle access to Earth. Earlier work often used only the longitudinal offset between the solar source and the Parker-spiral footpoint, but for high-energy SEPs even modest latitudinal offsets can significantly weaken connectivity and suppress prompt, highly anisotropic onsets. Observations and modeling consistently show that the most intense, sharply beamed GLEs originate from regions that are both longitudinally and latitudinally close to Earth’s heliographic position \citep[e.g.,][]{Dalla2010,GopalswamyMakela2014,Zhang2025}.
To incorporate this broader geometry, \citet{BrunoRichardson2021} introduced a practical formulation in which the connection angle is defined as the great--circle (spherical) distance between the parent flare site (latitude $\alpha_{F}$, longitude $\beta_{F}$) and the Parker--spiral footpoint at Earth (latitude $\alpha_{E}$, longitude $\beta_{E}$):
\begin{equation}\label{eq:great-circle}
\delta = \arccos\!\left[\sin(\alpha_{E})\sin(\alpha_{F}) 
+ \cos(\alpha_{E})\cos(\alpha_{F})\cos(\beta_{E}-\beta_{F})\right].
\end{equation}
The footpoint coordinates are obtained from the contemporaneous solar--wind speed using a standard Parker--spiral model\footnote{See Appendix in \citet{BrunoRichardson2021}.}. Although this metric does not capture all aspects of the true heliospheric connectivity, it provides a consistent way to account for both longitudinal and latitudinal offsets when characterizing SEP-event spatial distribution.

The variance in $A_{d}$ in our sample is dominated by the longitudinal deviation, but 
this reflects a selection bias rather than a lack of physical relevance: GLEs tend to originate relatively close to Earth’s heliographic latitude, restricting the range of $|\alpha_{E}-\alpha_{F}|$ and suppressing its statistical influence despite its well‑established importance for relativistic SEP access. 

The resulting connection angles for the ten GLEs in this study are listed in Table~\ref{tab:GLEs}.
Uncertainties shown in the following plots were estimated by propagating a 100\,km\,s$^{-1}$ solar-wind uncertainty together with a 5$^{\circ}$ uncertainty in the footpoint coordinates. For GLE~45, which lacks direct solar-wind measurements, we adopted 650\,km\,s$^{-1}$ and derived error bars by recomputing the angle at 550 and 750\,km\,s$^{-1}$, following prior studies of the disturbed heliospheric conditions during the October 1989 sequence (Section~\ref{GLE45}).

\subsection{Anisotropy vs. angle}\label{Anisotropy vs Angle}
\begin{figure*}[!t]
\centering
\caption{Proton anisotropy as a function of the nominal magnetic connection angle, computed from the full pitch‑angle distribution (left) and after subtracting the back‑scattered component (right).}
\begin{tabular}{cc}
\includegraphics[width=0.475\linewidth]{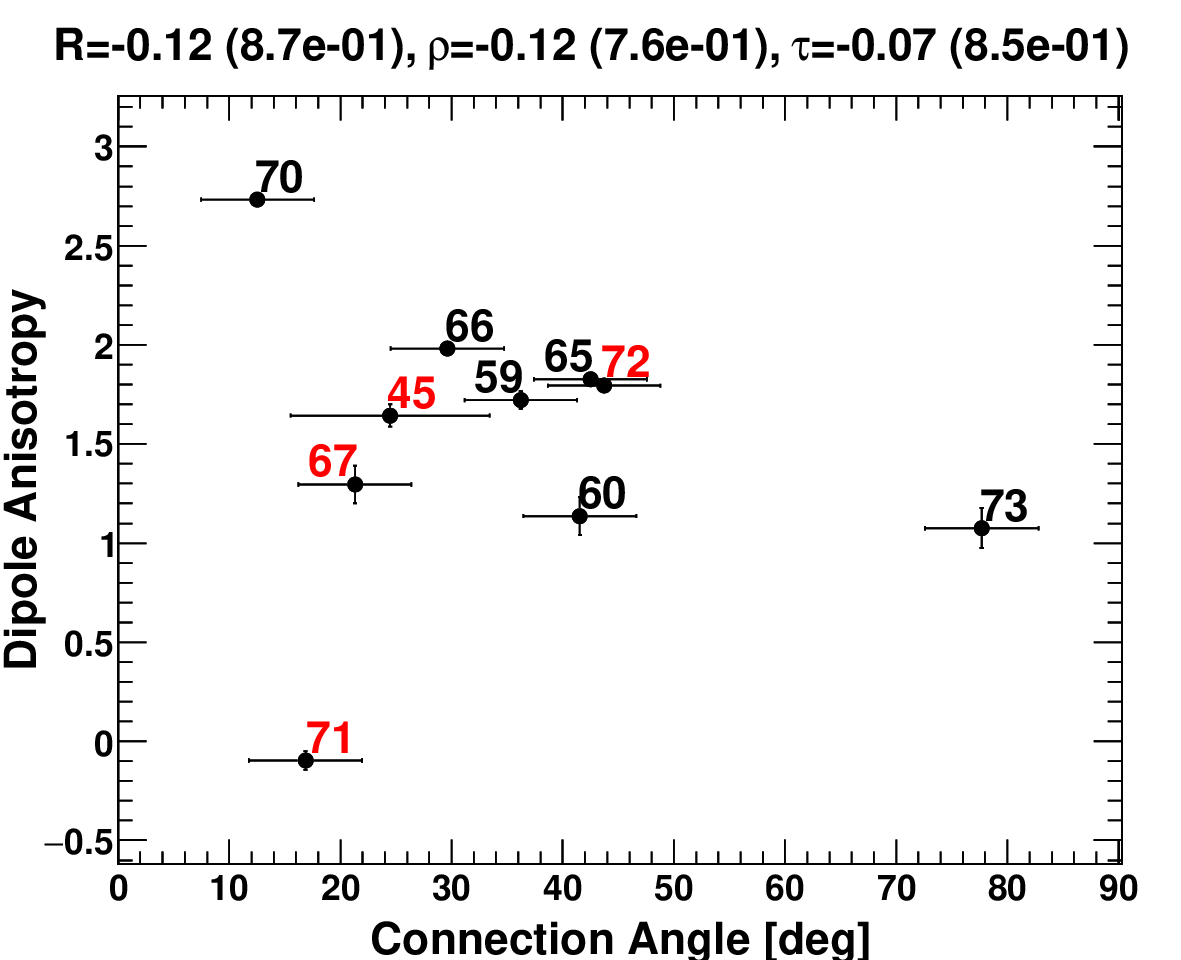} &
\includegraphics[width=0.475\linewidth]{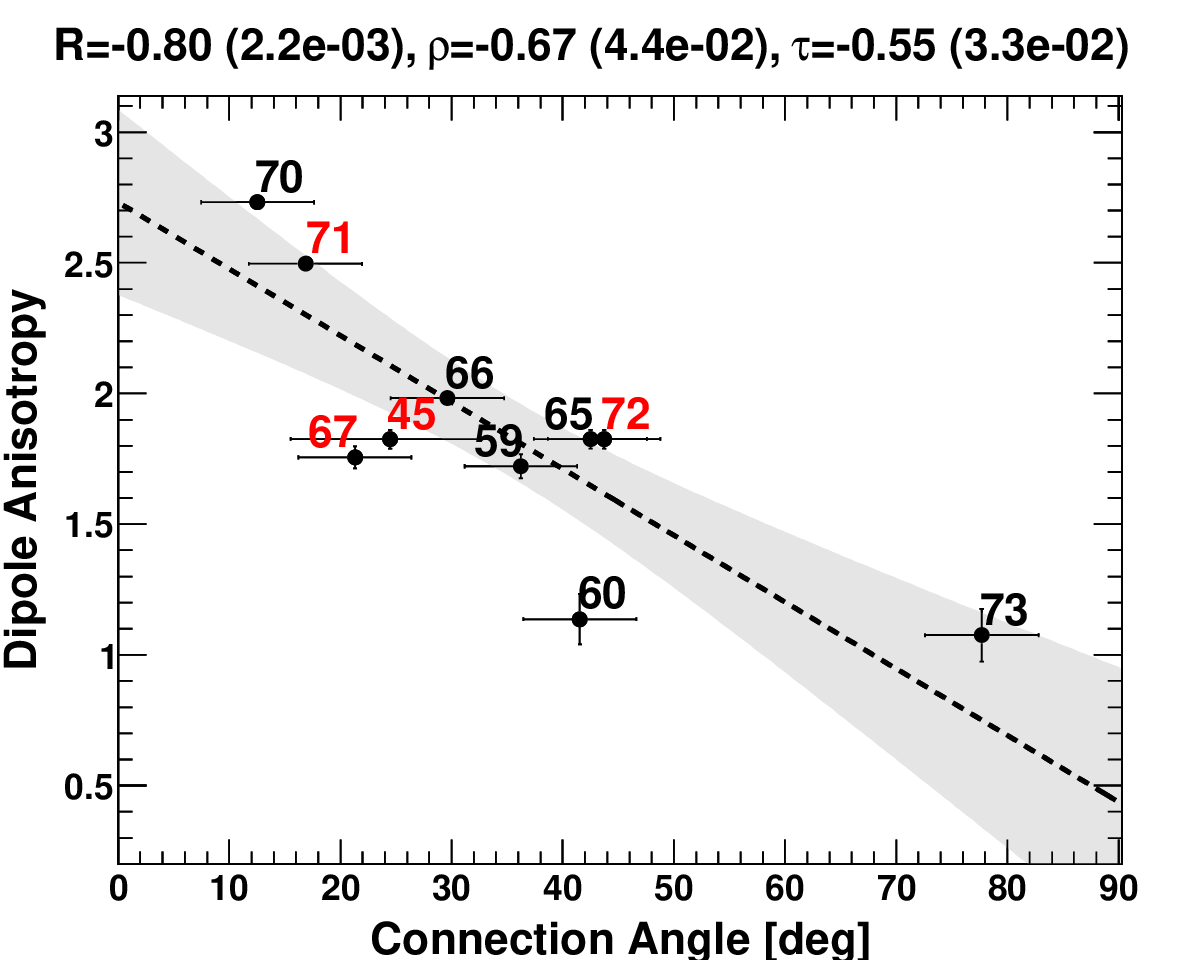} \\
\end{tabular}
\label{fig:Correlation_0}
\end{figure*}

Figure~\ref{fig:Correlation_0} presents the dipole anisotropy measured during the first five minutes of each GLE as a function of the nominal magnetic connection angle. The left panel displays results obtained from the full PAD, with events that include a measurable back-scattered component (GLEs \#45, \#67, \#71 and \#72) highlighted in red. The right panel shows the same comparison after subtraction of that back-scattered flux; the dashed line and the gray band denote the best‑fit trend and its associated 95\% confidence interval.
Correlation coefficients -- Pearson ($R$), Spearman ($\rho$), and Kendall ($\tau$) -- and associated p-values are reported on each panel to quantify linear and rank-based associations. Results obtained using the forward–inward anisotropy metric are qualitatively similar to those shown here for the dipole anisotropy.

The principal empirical finding is straightforward: when the back-scattered (sunward) component is present there is no systematic correlation between early anisotropy and connection angle, whereas subtraction of the back-scattered contribution reveals a statistically significant correlation. In other words, isolating the direct forward beam uncovers a clear connection-dependent behavior: smaller connection angles (better nominal connectivity) correspond to larger forward-beam anisotropy in the first five minutes of the event. This pattern is robust across anisotropy metrics and is reflected in the correlation statistics shown on the panels. The p-values 
associated with the correlation coefficients 
are all below conventional significance thresholds, indicating that the negative association observed after removal of the back-scattered component is unlikely to arise from random sampling variability.

\begin{figure}
\centering
\caption{Time evolution (0–60 min) of the Pearson correlation between anisotropy and magnetic connection angle. See text for the details.}
\includegraphics[width=0.95\linewidth]{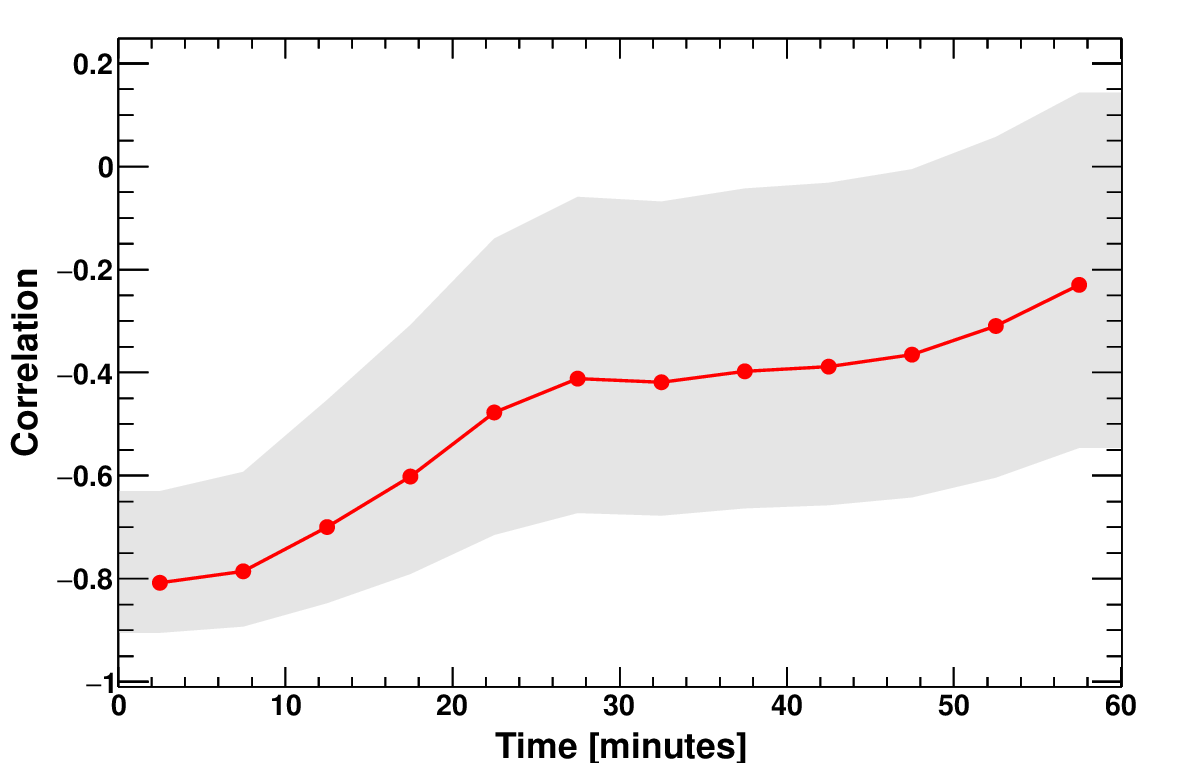} 
\label{fig:Correlation_vs_time}
\end{figure}

Physically, the observed correlation between connection angle and anisotropy is expected. Small connection angles map the observer to field lines that thread the acceleration region, enabling prompt, nearly scatter-free streaming and rapid isotropization at 1~AU. Larger connection angles imply that particles reaching the observer must either diffuse across field lines, sample shock flanks, or traverse longer, more convoluted trajectories; each of these increases the effective path length and the probability of encountering turbulent/sheath regions, which preserves a stronger field-aligned anisotropy during the early phase. 
The rigidity dependence of pitch-angle scattering further modulates this effect: lower-rigidity populations scatter more efficiently and will isotropize differently for the same connection geometry, so the observed angle–anisotropy relation is shaped by both geometry and the characteristic rigidities sampled.

The interpretation carries several practical and scientific implications. First, anisotropy—provided back-scattered or mirrored flux is removed—can serve as a quantitative diagnostic of nominal magnetic connectivity and early transport complexity. Second, the absence of a correlation in the full PAD demonstrates that reflected or return flux, which is highly sensitive to local magnetic geometry (ICME flux-rope orientation, mirror points, local curvature), can mask intrinsic connectivity signatures.
Third, because connection angle explains only part of the variance in early anisotropy, robust inference requires controlling for confounding factors such as characteristic rigidity, local turbulence/sheath metrics, and ICME preconditioning.

Although the anticorrelation between $A$ and connection angle is strong, the remaining spread in Figure~\ref{fig:Correlation_0} reflects the influence of additional acceleration and transport processes. Variations in shock geometry, seed populations, turbulence levels, field-line meandering, cross-field diffusion, and large-scale heliospheric structure—including the HCS—can all modulate the earliest PADs. Residual scatter may also arise from uncertainties and simplifying assumptions in the NM-based PAD reconstructions, such as yield-function systematics, geomagnetic tracing, and the neglect of rigidity and gyro-phase dependence.

The same qualitative relation between nominal connection angle and forward-beam anisotropy persists beyond the initial five-minute window, but its strength diminishes with time. Figure~\ref{fig:Correlation_vs_time} shows the Pearson correlation computed in successive time bins up to 60\,min after onset (back-scattered flux removed). The shaded band indicates the 1-\(\sigma\) confidence interval on the Pearson coefficient. A clear, statistically significant anticorrelation is present at early times and then decays steadily toward marginal significance over the first hour, indicating that the connectivity imprint is strongest during the prompt phase and is progressively erased by scattering, cross-field transport, and evolving shock–field geometry.

The temporal behavior is physically intuitive and diagnostically valuable. At onset the forward beam often propagates nearly scatter-free along well-connected field lines, so connection angle is a strong predictor of the early anisotropy imprint. As time progresses, however, pitch-angle scattering, cross-field transport in the sheath and magnetosheath, rigidity-dependent diffusion, and evolving shock–field geometry broaden and isotropize the distribution; these processes progressively erase the initial connectivity signature, producing the observed decay of the anticorrelation. Quantitatively, the early anticorrelation constrains the degree of initial magnetic connectivity, while the decay rate provides an empirical measure of effective scattering and perpendicular transport on 1-AU timescales and thus places bounds on parallel mean free paths and perpendicular diffusion coefficients.

\subsection{Isotropization vs. connection angle} \label{Isotropization vs Connection Angle}
A positive correlation between connection angle and isotropization time is physically plausible: small connection angles imply good nominal magnetic connectivity, allowing relativistic particles to stream nearly scatter-free and produce a prompt, field-aligned beam that typically relaxes rapidly at 1~AU, whereas large connection angles require cross-field access or sampling of shock flanks, increasing effective path length, exposure to turbulent/sheath regions, and the likelihood of delayed or secondary populations. These processes, together with the rigidity dependence of pitch-angle scattering (lower-rigidity particles scatter more efficiently), make a first-order positive angle–time trend intuitive.

\begin{figure*}[!t]
\centering
\caption{Isotropization time versus magnetic connection angle from fits to the full PAD (left) and to PADs with the back‑scattered component removed (right).}
\begin{tabular}{cc}
\includegraphics[width=0.475\linewidth]{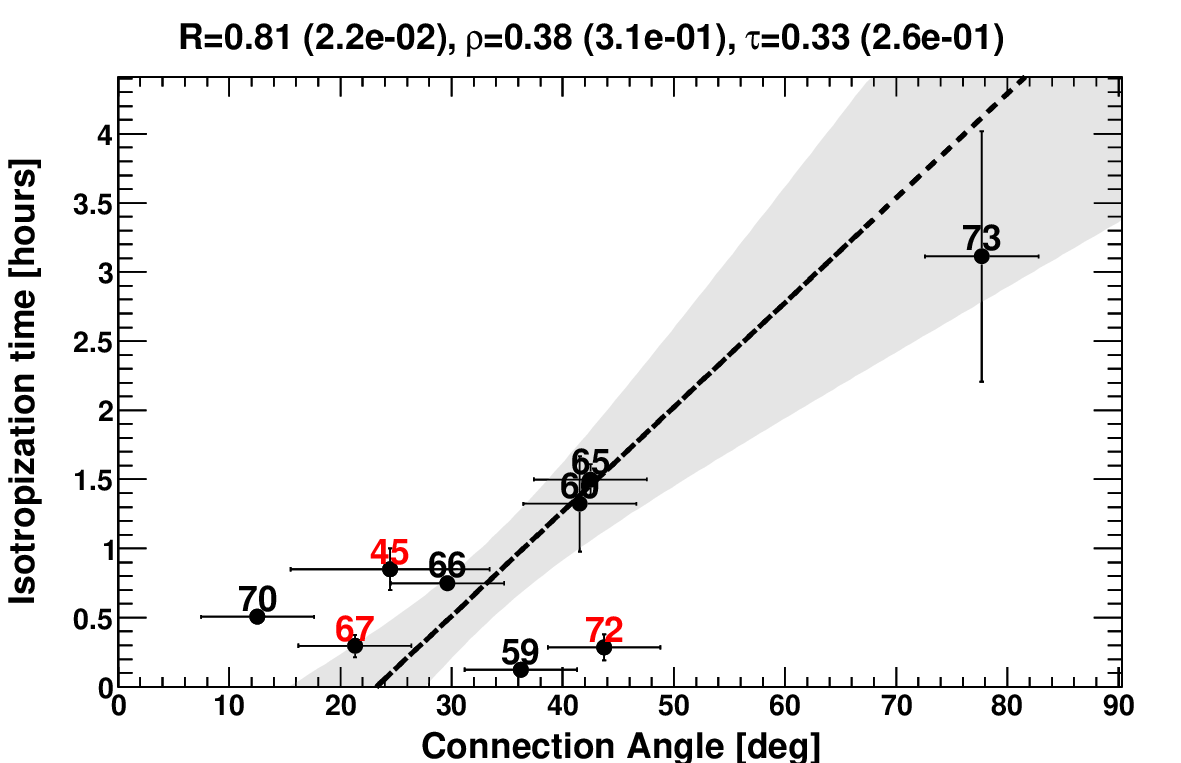} &
\includegraphics[width=0.475\linewidth]{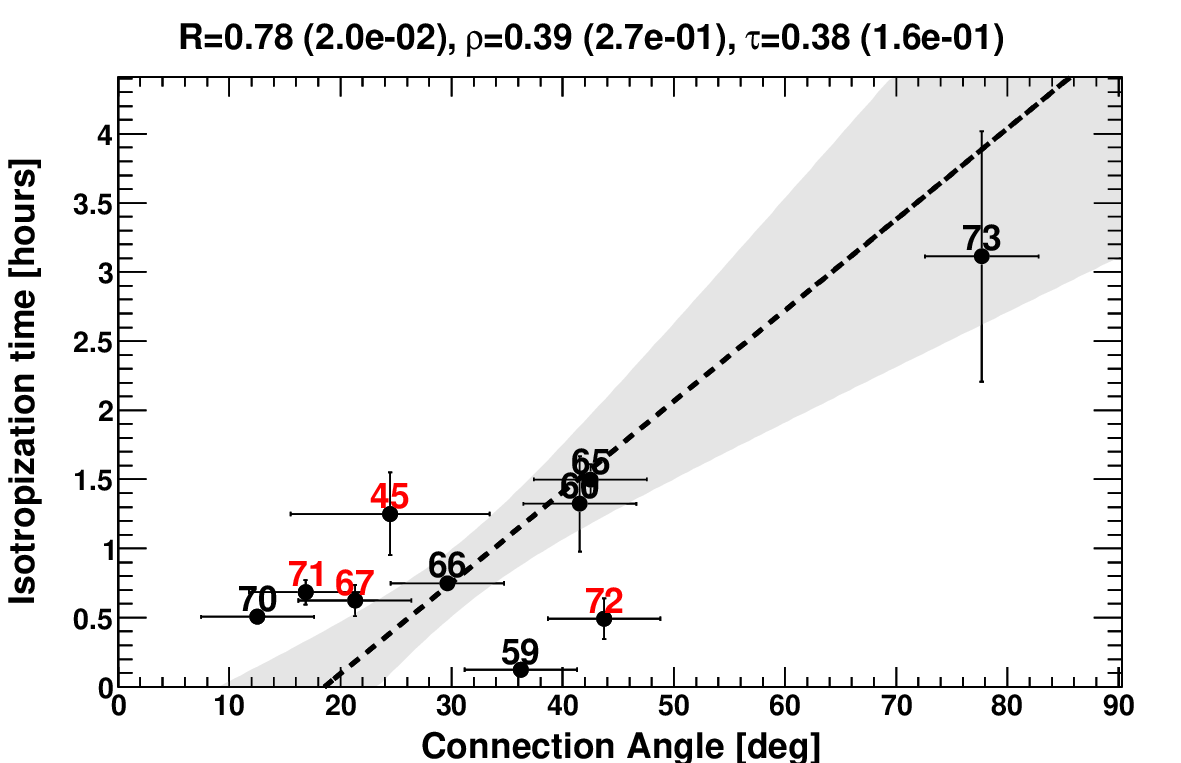} \\
\end{tabular}
\label{fig:Tau_vs_Angle}
\end{figure*}

Figure~\ref{fig:Tau_vs_Angle} tests this expectation. Using the full PAD (left panel), the plotted statistics indicate a moderate to strong linear association (\(R=0.78\), \(p=3.2\times10^{-2}\)), but weaker rank correlations (\(\rho=0.48\), \(p=0.19\); \(\tau=0.39\), \(p=0.18\)), implying that the apparent linear trend is influenced by a few high‑leverage events—most notably GLE~73. Observations in which a back-scattered contribution was identified are marked in red. The fitted regression line and its 95\% confidence interval (shaded area) suggest a systematic increase in isotropization time with connection angle, but the relatively high \(p\)-values for the rank‑based metrics caution against overinterpreting this trend as statistically robust across the full sample. However, the plotted points carry substantial uncertainties; error bars propagated from the PAD width materially increase the scatter and reduce the robustness of the fitted trend. After subtraction of the measurable back-scattered (sunward) component (right panel), the association is reduced (\(R=0.70\), \(p=4.4\times10^{-2}\); \(\rho=0.33\), \(p=0.35\); \(\tau=0.29\), \(p=0.29\)), indicating that the back-scattered population contributes disproportionately to the apparent angle--time relationship and that its removal reveals a weaker, less significant dependence. Note that when the back-scattered component is retained, GLE~71 is excluded because its PAD developed a rapidly evolving bidirectional structure that precluded a stable exponential fit; the absence of this point (a potential high‑leverage event) may therefore contribute to the stronger correlation observed in that case.

The lack of a significant correlation between isotropization time and connection angle does not contradict the strong anticorrelation between initial anisotropy and connection angle (Figure \ref{fig:Correlation_0}). The initial anisotropy primarily reflects source–observer magnetic connectivity and the fraction of prompt, field‑aligned particles that reach 1 AU with minimal scattering, whereas the isotropization time reflects the path‑integrated scattering, reflection, and mirroring experienced during transport. Large‑scale and transient heliospheric structures (ICMEs, sheaths, SIRs, and the HCS), local turbulence levels, and magnetospheric effects can strongly influence isotropization times without changing the nominal connection angle. Thus isotropization appears to be governed mainly by local transport conditions — for example, turbulence amplitude and spectral shape in sheaths and the magnetosheath, rigidity‑dependent scattering, the detailed shock–field geometry at the connected footpoint, and event‑to‑event preconditioning.

\subsection{Secondary correlations with flare class and CME speed}
Because faster CMEs tend to drive stronger and more extended shocks, and larger flares often accompany more energetic eruptions, one might expect more powerful events to produce stronger initial beaming (higher $A$) and, depending on whether the observed relaxation is dominated by scattering or by extended injection, either shorter or longer isotropization timescales. To test this, we computed Spearman rank correlations between the anisotropy metrics ($A$ and $\tau$) and the flare soft X-ray flux and CME speed listed in Table~\ref{tab:GLEs}. In contrast to the connection-angle trends discussed in Sections~\ref{Anisotropy vs Angle}--\ref{Isotropization vs Connection Angle}, the results show no coherent behavior. 
In practice, however, neither CME speed nor flare class is a reliable proxy for the shock properties or magnetic connectivity that control the earliest relativistic anisotropy. The limited event sample and the strong event-to-event variability in shock geometry, seed populations, and heliospheric conditions further weaken any apparent trends, leading to correlations that are small in magnitude and not always consistent in sign. As discussed in Section~\ref{Anisotropy vs Angle}, additional acceleration and transport processes—including variations in shock geometry, seed populations, turbulence levels, field-line meandering, cross-field diffusion, and large-scale heliospheric structure—introduce substantial scatter in the early PADs. These effects naturally obscure any simple dependence on CME speed or flare class, reinforcing that magnetic connection angle is the primary control on the initial anisotropy.

\section{Conclusions}

We have presented a uniform, event‑resolved analysis of the early PADs for ten well‑observed GLE events and have found a robust, monotonic decline of initial anisotropy with increasing magnetic connection angle: events with small connection angles exhibit strong, more persistent forward‑directed beams, while poorly connected events show systematically weaker with and more rapidly decaying anisotropies. This empirical relation holds across a wide range of flare classes and CME speeds, indicating that magnetic connectivity and interplanetary transport, rather than eruption magnitude alone, are the primary controls on the directional properties of the earliest relativistic arrivals at Earth.

The conclusion is supported by three complementary methodological advances. First, we used consistently reconstructed NM‑based PADs and time‑resolved spectral fits to ensure comparability across events. Second, PCA separates coupled spectral–angular evolution from independent transport signatures, clarifying which changes are source driven and which reflect propagation. Third, by explicitly identifying and removing secondary sunward (back‑scattered) components from the PAD fits, we isolate the intrinsic relaxation of the primary forward beam and show that many departures from simple exponential decay are attributable to reflected or delayed populations rather than prolonged source injection.

Operationally, rapid footpoint connectivity estimates can be used as a timely prior for the expected beam direction by assuming the initial anisotropy is aligned with the local IMF; because full, global PAD reconstructions are not available in true real time, this geometry‑based prior provides immediate, actionable guidance for directional radiation‑risk assessments and instrument‑pointing decisions and can be updated as NM‑based PAD inversions and improved mapping become available.

We emphasize several important caveats. The analysis is based on a curated sample of ten well‑observed GLEs, and the limited sample size and selection toward large eruptions caution against unqualified extrapolation to the broader SEP population. Systematic uncertainties in NM‑based reconstructions (yield‑function calibration, geomagnetic tracing, PAD functional form, and implicit rigidity/gyro‑phase assumptions) can affect quantitative anisotropy metrics and contribute to the residual scatter. Physically, a suite of acceleration and transport processes—shock obliquity and formation height, injection physics and seed populations, time dependence and spatial extent of the accelerator, localized acceleration structures, stochastic/wave‑driven acceleration, turbulence, adiabatic focusing, field‑line meandering, cross‑field diffusion, magnetic mirroring, and large‑scale heliospheric structures such as the HCS and CIRs—introduce event‑to‑event variability that is not captured by simple correlations with CME speed or flare class.

These results provide an event‑resolved, quantitative benchmark for focused‑transport and shock‑acceleration models. To advance understanding and test the generality of the empirical relation reported here, we recommend targeted modeling that couples realistic shock geometry and time‑dependent injection with focused‑transport simulations, coordinated high‑rigidity point measurements to improve NM yield‑function anchoring, and continued refinement of NM calibrations.

\begin{acknowledgements}
We thank Alexander Mishev for providing support with the neutron‑monitor angular and energy distribution calculations.
A.~B. acknowledges support from NASA Heliophysics Space Weather Research Program's CLEAR SWx Center of Excellence award 80NSSC23M0191, and from NASA program 80GSFC24M0006. S.~D. acknowledges support from the UK STFC (grants ST/V000934/1 and ST/Y002725/1).
\end{acknowledgements}

\bibliographystyle{aa}
\bibliography{bibliography}

\end{document}